\pdfoutput=1
\documentclass[conference]{IEEEtran}

\usepackage{amsthm} 

\newtheorem{axiom}{Axiom}

\usepackage{supertabular,array}
\usepackage{hyphenat} 
\usepackage{multirow}
\usepackage[hide]{todo}

\usepackage{cite}
\usepackage[cmex10]{amsmath} 
  \usepackage[pdftex]{graphicx}
\usepackage[caption=false,font=footnotesize]{subfig}
\usepackage{fixltx2e}

\newcommand{\paperTitle}{Towards a Synergy-based Approach to Measuring Information Modification}
\newcommand{\theKeywords}{information modification, information storage, information transfer, intrinsic computation, complex systems, information theory, cellular automata, particle collisions}

    \usepackage{hyperref}
    \hypersetup{
        pdfauthor = {Joseph T. Lizier, Benjamin Flecker and Paul L. Williams},
        pdftitle = {\paperTitle},
        pdfsubject = {Version 2.1, SVN 245},
        pdfkeywords = {\theKeywords},
        pdfcreator = {LaTeX with hyperref package},
        pdfproducer = {pdflatex}}

\newcommand{\fig}[1]{Fig.~\ref{fig:#1}}
\newcommand{\eq}[1]{Eq.~(\ref{eq:#1})}

\newcommand{\secRef}[1]{Section \ref{sec:#1}}

\newcommand{\tableRef}[1]{Table \ref{table:#1}}

\DeclareMathOperator*{\argmin}{arg\,min} 

\begin{document}


\title{\paperTitle}

\author{\IEEEauthorblockN{Joseph T. Lizier\IEEEauthorrefmark{1}\IEEEauthorrefmark{2},
Benjamin Flecker\IEEEauthorrefmark{3},
Paul L. Williams\IEEEauthorrefmark{4}}
\IEEEauthorblockA{\IEEEauthorrefmark{1}CSIRO Information and Communications Technology Centre, PO Box 76, Epping, NSW 1710, Australia}
\IEEEauthorblockA{\IEEEauthorrefmark{2}Max Planck Institute for Mathematics in the Sciences,
Inselstra{\ss}e 22, 04103 Leipzig, Germany}
\IEEEauthorblockA{\IEEEauthorrefmark{3}Department of Physics, Indiana University, Bloomington, IN 47405-7105, USA}
\IEEEauthorblockA{\IEEEauthorrefmark{4}Cognitive Science Program, Indiana University, Bloomington, IN 47406, USA}
\newline Email: joseph.lizier@csiro.au}

\maketitle

\begin{abstract}
Distributed computation in artificial life and complex systems is often described in terms of component operations on information: information storage, transfer and modification.
Information modification remains poorly described however, with the popularly-understood examples of glider and particle collisions in cellular automata being only quantitatively identified to date using a heuristic (separable information) rather than a proper information-theoretic measure.
We outline how a recently-introduced axiomatic framework for measuring information redundancy and synergy, called partial information decomposition, can be applied to a perspective of distributed computation in order to quantify component operations on information.
Using this framework, we propose a new measure of information modification that captures the intuitive understanding of information modification events as those involving interactions between two or more information sources.
We also consider how the local dynamics of information modification in space and time could be measured, and suggest a new axiom that redundancy measures would need to meet in order to make such local measurements.
Finally, we evaluate the potential for existing redundancy measures to meet this localizability axiom.
\end{abstract}


\IEEEpeerreviewmaketitle

\section{Introduction}
\label{sec:intro}

Considering how variables are dynamically composed of information from various sources is a topical subject in physics, complex systems and artificial life.
For example, we have seen the dynamics of information studied in cellular automata \cite{sha06,liz08a,liz10e,liz12a}, brain-body-environment systems \cite{will10b}, financial systems \cite{har09a}, models of gene regulatory networks \cite{liz11b}, and the relation of network structure to these dynamics \cite{liz12b}.

There are several perspectives on how the composition or ``credit assignment'' of information could be studied (e.g. \cite{lud08,kah09,will10a,liz13a}).
We study \emph{information dynamics} through the lens of \emph{distributed computation}, focussing on operations of \emph{information storage}, \emph{transfer} and \emph{modification} \cite{liz08a,liz10e,liz12a,liz13a} (described in \secRef{infoDynamics}).
This is because these terms are generally well-understood (e.g. information transfer as directed coupling between two nodes) especially in comparison to general notions of complexity, and can be measured on any type of time-series data.
Furthermore, computation is the language in which dynamics in complex systems are often described (e.g. Langton's ``Computation at the edge of chaos'' \cite{lang90}).

Crucially, this approach has provided key theoretic insights into cellular automata (CAs), a critical proving ground for any theory on the fundamental nature of distributed computation in complex systems.
CAs are discrete dynamical systems with an array of cells that synchronously update their state as a function of a fixed number of spatial neighbors cells using a uniform rule \cite{wolf02}.
Elementary CAs (ECAs) are 1D arrays of binary state cells with one neighbor on either side.
Studies of computation in CAs typically focus on emergent structures, such as domains, particles, and gliders.
A \emph{domain} is a set of background configurations, any of which will update to another such configuration in the absence of disturbances. 
\emph{Particles} are dynamic, coherent spatiotemporal structures against this background: \emph{gliders} are regular particles, and \emph{blinkers} are stationary gliders.
The information dynamics approach provided the first quantitative evidence \cite{liz08a,liz10e,liz12a} for the conjecture \cite{lang90} that blinkers are information storage entities, that particles are associated with information transfer, and that particle collisions correspond to information modification events.

Despite the success of this perspective, we do not have a complete quantitative understanding of the notion of information modification.
It is often colloquially described as the processing of information into a new form.
It has been viewed as a pivotal operation for biological neural networks and models thereof \cite{kin06a,at92,san02},
where it has been suggested as a potential biological driver \cite{at92}. It is also a key operation in collision-based computing \cite{adama02}.
As such, information modification operations are likely to be required to support complex behavior in artificial life and biological systems.

To be specific, information modification has been interpreted to mean interactions between transmitted and/or stored information which result in a modification of one or the other \cite{lang90}.
This interpretation specifically juxtaposes modification against storage and transfer, viewing it as a dynamic combination or synthesis of information from different sources. Modification therefore involves a non-trivial processing of information from two or more (storage or transfer) sources, rather than a trivial retrieval, movement or translation of one source of information alone.
The separable information was introduced previously to study information modification \cite{liz10e}.
Whilst it quantitatively identified particle collisions in cellular automata as modification events, the separable information is a heuristic rather than a proper information-theoretic measure.

Much recent attention \cite{will10a,will11a,fleck11a,grif12a,hard12a,bert12a,tim12a} has been focused on information-theoretic measures of redundancy and synergy between information sources in creating outcomes in a target or destination variable.
These efforts began with the abstract, axiomatic \emph{partial information decomposition} (PID) framework of Williams and Beer \cite{will10a}, as described in \secRef{pid}.
The concept of synergy, as formalized in the PID framework, is particularly appealing for the notion of information modification described above, as it explicitly quantifies the information associated with two or more information sources that is not present in any subset of those sources.
In \secRef{infoModification}, we propose a measure of information modification based on the PID framework and its concept of synergy, and discuss its merits relative to previously proposed measures of information modification.  In particular, we argue that (1) our measure clarifies the intertwined nature of information modification and transfer---with modification corresponding to the synergistic parts of transfer---and (2) our measure has the desirable property that modification events of various orders can be hierarchically decomposed into separately quantifiable terms.

Furthermore, we describe in \secRef{localising} how, in order to study the \emph{dynamics} of such modification on a \emph{local} scale in space and time, we require the concrete measures of redundancy and synergy applied via the PID framework to be localizable themselves.
We define a new axiom for such concrete measures to satisfy in terms of localizability, but show that $I_{\min}$ \cite{will10a} (the most prominent redundancy measure) does not satisfy it.
Finally, we consider the future prospects for a concrete measure that could be applied to properly quantify information modification on a local scale in space and time.

\section{Information theory}
\label{sec:infotheory}

In this section, we briefly introduce two key background concepts from information theory \cite{shan48,cover91,mac03} related to our analysis: the nature of redundant and synergistic contributions of two variables to the information in another, and the local value of information measures at specific observations.

%

The \emph{mutual information} (MI) between $X$ and $Y$ measures the average reduction in uncertainty about $x$ that results from learning the value of $y$, or vice versa: $I(X;Y) = H(X)-H(X|Y)$, where $H(X) = -\sum_{x} p(x) \log_2{p(x)}$ and $H(X|Y) = -\sum_{x,y} p(x,y) \log_2{p(x|y)}$ are the \emph{Shannon entropy} and \emph{conditional entropy} respectively.
The \emph{conditional mutual information} between $X$ and $Y$ given $Z$ is the MI between $X$ and $Y$ when $Z$ is known: $I(X;Y|Z) = H(X|Z)-H(X|Y,Z)$.
One can consider the MI from two variables $Y_1, Y_2$ jointly to another, $I(X; Y_1,Y_2)$, and decompose this into the information carried by the first variable plus that carried by the second conditioned on the first: $I(X; Y_1,Y_2) = I(X; Y_1) + I(X; Y_2 | Y_1)$.
It is crucial to understand that a conditional MI $I(X;Y|Z)$ may be either larger or smaller than the related unconditioned MI $I(X;Y)$ \cite{mac03}; the conditioning removes information \emph{redundantly} held by the source $Y$ and the conditioned variable $Z$ about $X$, but also includes \emph{synergistic} information about $X$ which can only be decoded with knowledge of both the source $Y$ and conditioned variable $Z$.
These components \emph{cannot} be teased apart with traditional information-theoretic analysis; the partial information decomposition (\secRef{pid}) was introduced for this purpose \cite{will10a}.

Next, note that the aforementioned information-theoretic quantities are \emph{averages} over all of the observations used to compute the relevant probability distribution functions (PDFs). One can also write down \emph{local} or pointwise measures for each of these quantities, representing their value for one specific observation or configuration of the variables $(x,y,z)$ being observed. The average of a local quantity over all observations is of course the relevant average information-theoretic measure.
Applied to time-series data, local measures tell us about the \textit{dynamics} of information in the system, since they vary with the specific observations in time, and local values are known to reveal more details about the system than the averages alone \cite{sha01a,sha06}.
For example, the local mutual information \cite{fano61} $I(X = x; Y = y) = i(x;y) = \log_2{p(x \mid y) / p(x)}$ for a specific observation $(x,y)$ is the information held in common between the specific values $x$ and $y$.
(By convention, we use lower case symbols for the local quantities.)
Indeed, the form of $i(x;y)$ is derived directly from four postulates \cite[ch. 2]{fano61}: once-differentiability, similar form for conditional MI, additivity (i.e. $i(\left\{y_n,z_n\right\};x_n) = i( y_n ;x_n) + i(z_n; x_n \mid y_n)$), and separation for independent ensembles.
This derivation also means that $i(x;y)$ is uniquely specified, up to the base of the logarithm.
Of course, $I(X;Y) = \left\langle i(x;y) \right\rangle$, and like $I(X;Y)$, $i(x;y)$ is symmetric in $x$ and $y$ (see further discussion in \cite{liz10a}).
Importantly, $i(x;y)$ may be positive or negative, meaning that one variable can either positively inform us or actually \emph{misinform} us about the other. An observer is misinformed where, conditioned on the value of $y$ the observed outcome of $x$ was \emph{relatively} unlikely as compared to the unconditioned probability of that outcome (i.e. $p(x | y) < p(x)$).

\section{Information dynamics}
\label{sec:infoDynamics}

A local framework for \emph{information dynamics} has recently been introduced in \cite{liz07b,liz08a,liz10e,liz12a,liz13a}.
This framework examines how the next value $x_{n+1}$ of a destination variable is \textit{computed} in terms of how much of that information came from the past state of that variable (\emph{information storage}), how much came from respective source variables (\emph{information transfer}), and how those information sources were combined (\emph{information modification}).
The framework has a particular focus on the \emph{dynamics} of these operations in time and space, and so provides spatiotemporal \emph{information profiles} for each measure.
In this section, we describe how the framework measures information storage and transfer, before considering information modification in \secRef{infoModification}.

\subsection{Information storage}

\emph{Information storage} is the amount of information from the past of a process that is relevant to or will be used at some point in its future.
In terms of the \emph{dynamics} of information processing, we focus on how much of the stored information is actually \emph{in use} in computing the current value of the process.
As such, the \emph{active information storage} (AIS) $A_X$ was introduced \cite{liz12a} to explicitly measure how much of the information from the past of a process $X$ is observed to be \emph{in use} in computing its next value.
$A_X$ is the average MI between realizations $\mathbf{x^{(k)}}_{n}  = \left\{ x_{n-k+1}, \ldots , x_{n-1}, x_n \right\}$ of the past \emph{state} $\mathbf{X^{(k)}}$ and the corresponding realizations $x_{n+1}$ of the \emph{next value} $X'$ of a given time series $X$:
\begin{align}
	A_X(k) = I ( \mathbf{X^{(k)}};X' )
		\label{eq:activeStorageEstimate}.
\end{align}
We require $\lim_{k \rightarrow \infty}$ in general, unless $x_{n+1}$ is conditionally independent of the far past values $\mathbf{x^{(\infty)}}_{n-k}$ given $\mathbf{x^{(k)}}_{n}$ \cite{liz12a}.

We can then extract the \emph{local} active information storage $a_X(n+1)$ \cite{liz12a} as the amount of information storage attributed to the specific configuration or realization $(\mathbf{x^{(k)}}_n,x_{n+1})$ at time step $n+1$; i.e. the storage in use by the process at $n+1$:
\begin{align}
	A_X(k) & = \left\langle a_X(n+1,k) \right\rangle_n
		\label{eq:averageActiveFinite}, \\
	a_X(n+1,k) & = i(\mathbf{x^{(k)}}_n ; x_{n+1}) = \log_{2}{\frac{p(\mathbf{x^{(k)}}_n,x_{n+1})}{p(\mathbf{x^{(k)}}_n)p(x_{n+1})}}
	  \label{eq:activeLocalMemory}.
\end{align}

As a local MI, $a_X(n+1,k)$ may be positive or negative, meaning the past history of the variable can either positively inform us or actually \emph{misinform} us about its next state. 

\begin{figure}[!t] 
  \begin{center}
    \subfloat[Raw CA]{\label{fig:54CA}\makebox[0.45\columnwidth]{\includegraphics[trim= 45 33 110 31,clip=true,height=0.43\columnwidth]{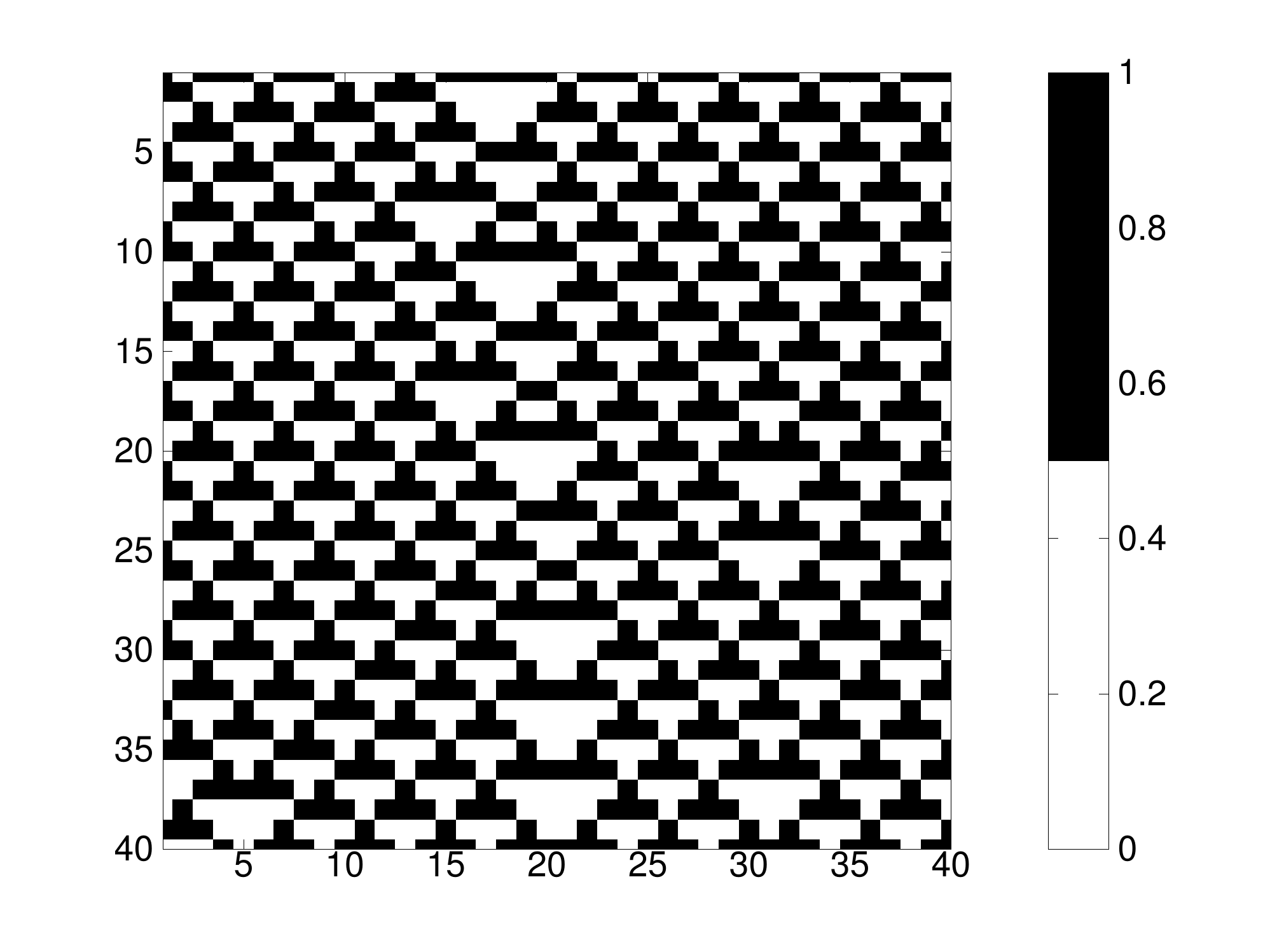}}}
    \subfloat[$a(i,n,k=16)$]{\label{fig:54piOrder1}\makebox[0.55\columnwidth]{\includegraphics[trim= 45 33 45 31,clip=true,height=0.43\columnwidth]{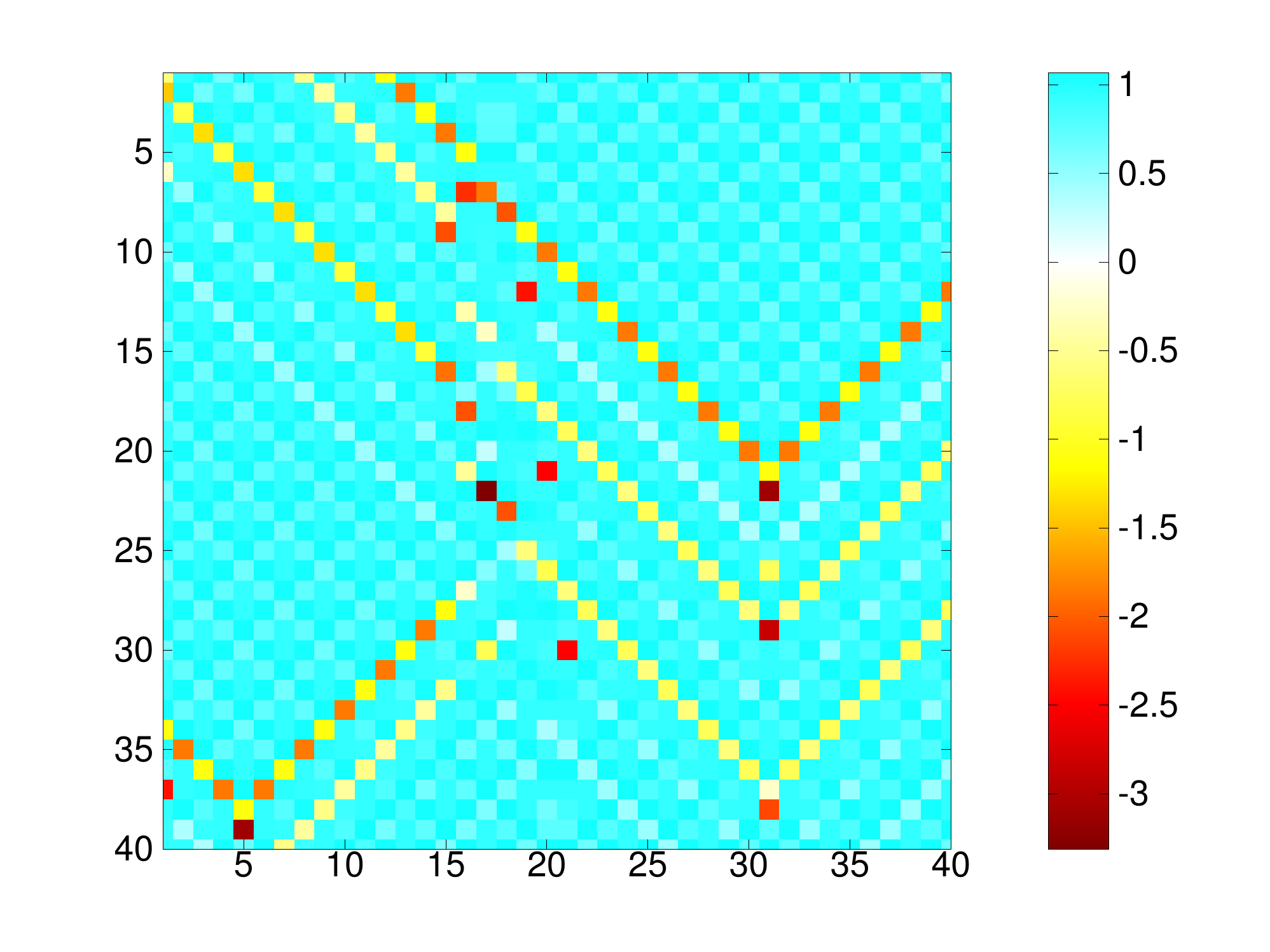}}} 
    \caption[Local profile of AIS]{Local profile of AIS \ensuremath{a(i,n,k=16)} in bits for each cell \ensuremath{i} for each time step \ensuremath{n} in \subref{fig:54piOrder1} for the raw states of CA rule 54 in \subref{fig:54CA}.}
    \label{fig:caLocalResults}
  \end{center}
\end{figure}

As reported in \cite{liz12a} (with sample results in \fig{caLocalResults}),
when applied to CAs the local AIS takes on large positive values in the domain and blinkers, since for these entities the next state is predictable from the destination's past.
This was the first direct quantitative evidence that blinkers and domains were the dominant information storage entities in CAs.
Furthermore, negative values are measured when gliders are encountered, because the past of the destination (being in the domain) would misinformatively predict domain continuation.

\subsection{Information transfer}

Information transfer is defined as the amount of information that a source provides about a destination's next state in the context of the destination's past. This definition pertains to Schreiber's transfer entropy (TE) measure \cite{schr00}.
The TE captures the average MI from realizations $y_{n}$ of a source\footnote{TE can consider realizations of the source \emph{state} $y^{(l)}_n$. This is appropriate where the observations $y$ mask a hidden causal process to $X$, or where multiple past values of $Y$ in addition to $y_n$ are causal to $x_{n+1}$ \cite{liz10a}.} $Y$ to the corresponding realizations $x_{n+1}$ of the destination $X'$, conditioned on realizations $\mathbf{x^{(k)}}_{n}$ of the destination's previous state $\mathbf{X^{(k)}}$:
\begin{equation}
	T_{Y \rightarrow X}(k) = I ( Y; X' \mid \mathbf{X^{(k)}} )
	\label{eq:te}.
\end{equation}
%
Different values of $k$ produce different results here, but
in alignment with $A_X(k)$, in general one should take the limit $k \rightarrow \infty$ here (except for similar conditional independence cases),
in order to properly interpret the transfer entropy as a measure of information transfer \cite{liz08a,liz10a}.

We can then extract the \emph{local transfer entropy} $t_{Y \rightarrow X}(n+1)$ \cite{liz08a} as the transfer attributed to the specific realization $(x_{n+1},\mathbf{x^{(k)}}_{n},y_{n})$ at time step $n+1$; i.e. the amount of information transfered from $Y$ to $X$ at $n+1$:
\begin{align}
	T_{Y \rightarrow X}(k) & = \left\langle t_{Y \rightarrow X}(n+1,k) \right\rangle
		\label{eq:teExpectationFiniteK}, \\
	t_{Y \rightarrow X}(n+1, k) & = \log_{2}{ \frac{ p(x_{n+1} \mid \mathbf{x^{(k)}}_{n},y_{n})}{p(x_{n+1} \mid \mathbf{x^{(k)}}_{n})}}
		\label{eq:localTE_genericLimit}, \\
		& = i(y_{n}; x_{n+1} \mid \mathbf{x^{(k)}}_{n})
		\label{eq:localTE_conditionalMI}.
\end{align}
%
For proper interpretation as information transfer, $Y$ is constrained among the $g$ causal information contributors to $X$, say $Y \in \{ Y_1, \ldots , Y_g \} \setminus X$ \cite{liz10a}.
Importantly, the information conditioned on by the TE
is that provided by the AIS.

Like local MI, local TE may be either positive or negative.
As reported in \cite{liz08a}, when applied to CAs it is typically strongly positive at gliders when measured in the same direction as the glider's motion (e.g. information transfer across one cell to the right per unit time).
Note: this result only holds for large $k$, i.e. when storage and transfer are properly separated.
These results provided the first quantitative evidence for the long-held conjecture that particles are the dominant information transfer entities in CAs.
Negative values imply that the source \emph{misinforms} an observer about the next state of the destination in the context of the destination's past, and are typically found when TE is measured orthogonally to a moving glider.

TE can also be conditioned on other possible sources $Z$ to account for their effects on the destination. The \emph{conditional} transfer entropy was introduced for this purpose \cite{liz08a,liz10e}:
\begin{align}
	T_{Y \rightarrow X \mid Z}(k) & = I ( Y; X' \mid \mathbf{X^{(k)}}, Z )
	\label{eq:teCondEstimate}, \\
	T_{Y \rightarrow X \mid Z}(k) & = \left\langle t_{Y \rightarrow X 
	\mid Z}(n+1,k) \right\rangle
		\label{eq:teCondExpectationFiniteK}, \\
	t_{Y \rightarrow X \mid Z}(n+1, k) & = \log_{2}{ \frac{ p(x_{n+1} \mid \mathbf{x^{(k)}}_{n},y_{n},z_n)}{p(x_{n+1} \mid \mathbf{x^{(k)}}_{n}, z_{n})}}
		\label{eq:localTE_genericLimit}, \\
		& = i(y_{n}; x_{n+1} \mid \mathbf{x^{(k)}}_{n}, z_n)
		\label{eq:localTE_conditionalMI}.
\end{align}

We specifically refer to the conditional TE as the \emph{complete transfer entropy} ($T^c_{Y \rightarrow X}(k)$ and $t^c_{Y \rightarrow X}(n+1, k)$) when it conditions on \emph{all} other causal sources $Z$ to the destination $X$ \cite{liz08a}.
For clarity then, we refer to $T_{Y \rightarrow X}$ simply as the \emph{apparent} transfer entropy \cite{liz08a}.
As conditional MI terms, these TEs may be larger or smaller than the unconditioned MIs;
we consider how such redundancies and synergies can be specifically measured in the next section.

Finally, note that one can decompose the MI from the sources to destination as a sum of incrementally conditioned MI terms \cite{liz10e,liz10a}; e.g. for a two source system:
\begin{align}
	I(X';\mathbf{X^{(k)}},Y_1,Y_2) & = I(X';\mathbf{X^{(k)}}) + I(X';Y_1 \mid \mathbf{X^{(k)}}) +  \nonumber\\
	& \ \ \ + I(X';Y_2 \mid \mathbf{X^{(k)}},Y_1)
	\label{eq:miBreakdown1}, \\
	& = A_X + T_{Y_1 \rightarrow X}(k) + T_{Y_2 \rightarrow X \mid Y_1}(k) 
	\nonumber.	
\end{align}
This equation could be reversed in the order of $Y_1$ and $Y_2$, and its correctness is independent of $k$ (so long as $k$ is large enough to capture the causal sources in the past of the destination).

\section{Partial information decomposition}
\label{sec:pid}
\subsection{Abstract definition}
\label{sec:pidAbstract}

The PID framework provides a general method of decomposing the information $I(X;\mathbf{A})$ that a set of source variables $\mathbf{A} = \{\mathbf{A}_1, \ldots, \mathbf{A}_r\}$ provide about a destination variable $X$ \cite{will10a}.
The core idea underlying this method is a measure of \emph{redundancy}  $I_\cap(X; \mathbf{A}_1, \ldots, \mathbf{A}_r)$, which captures the overlapping information that sources $\mathbf{A}_1, \ldots, \mathbf{A}_r \subseteq \mathbf{A}$ (which may be joint variables in general) share about the destination $X$.
Intuitively, redundancy acts on information sources like the intersection operator acts on sets, capturing the information that is common to all sources.
Indeed, the redundancy measure $I_\cap$ is defined by the following axioms, each of which is analogous to a basic property of set intersection:
\begin{axiom}
\emph{Symmetry}: $I_\cap$ is symmetric in the $\mathbf{A}_i$'s.
\end{axiom}
\begin{axiom}
\emph{Self-redundancy}: $I_\cap(X;\mathbf{A}_i) = I(X;\mathbf{A}_i)$.
\end{axiom}
\begin{axiom}
\emph{Monotonicity}: $I_\cap(X; \mathbf{A}_1, \ldots, \mathbf{A}_{r-1}, \mathbf{A}_r ) \leq I_\cap(X; \mathbf{A}_1, \ldots, \mathbf{A}_{r-1} )$ with equality if $\mathbf{A}_{r-1} \subseteq \mathbf{A}_r$.
\end{axiom}

\begin{figure}[t]
	\begin{center}
		\includegraphics[width=0.32\textwidth]{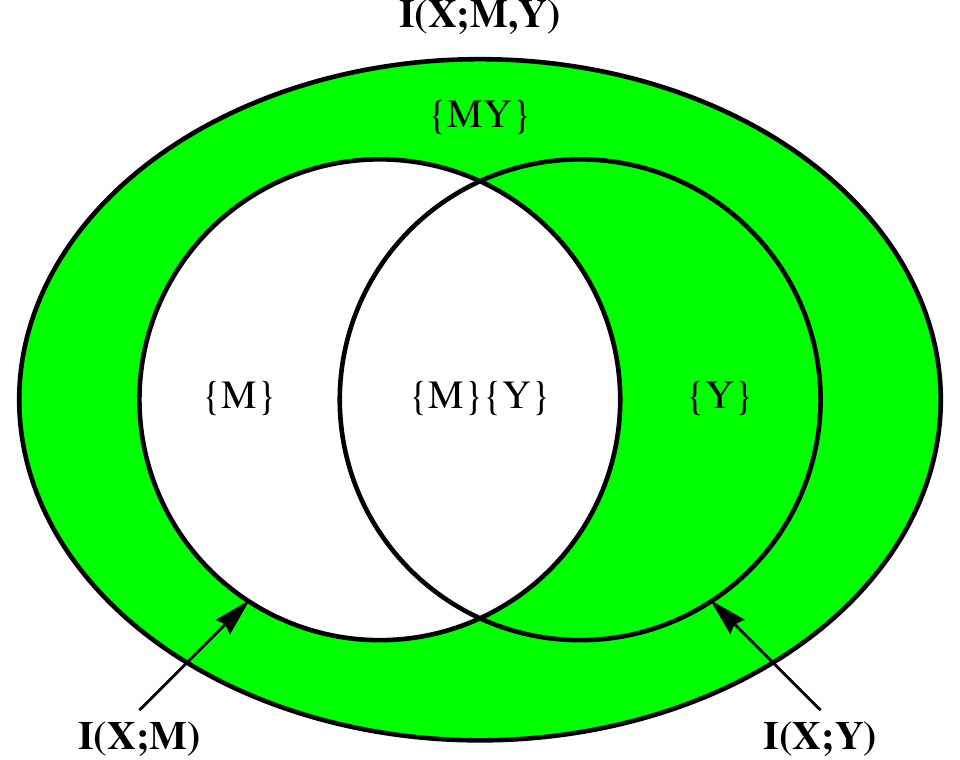}
	\end{center}
	\caption{\label{fig:partial2Source}Partial information diagram of information $I(X; M,Y)$ in $X$ from \emph{two} source variables $M, Y$ (ignoring the colors).
$\{M\}\{Y\}$ represents the redundant information in the two sources, $\{M\}$ and $\{Y\}$ represent the unique information from each source, and $\{M,Y\}$ represents the synergistic information from the sources.
If we consider $M$ to be the past state $\mathbf{X^{(k)}}$ of the destination $X$, and $Y$ as another causal source, then this PI-diagram partitions the AIS (white) and TE (green). (This is called the PI-diagram for \emph{three} variables in \cite{will10a}, including the destination variable.)}
\end{figure}

\begin{figure}[t]
	\begin{center}
		\includegraphics[width=0.48\textwidth]{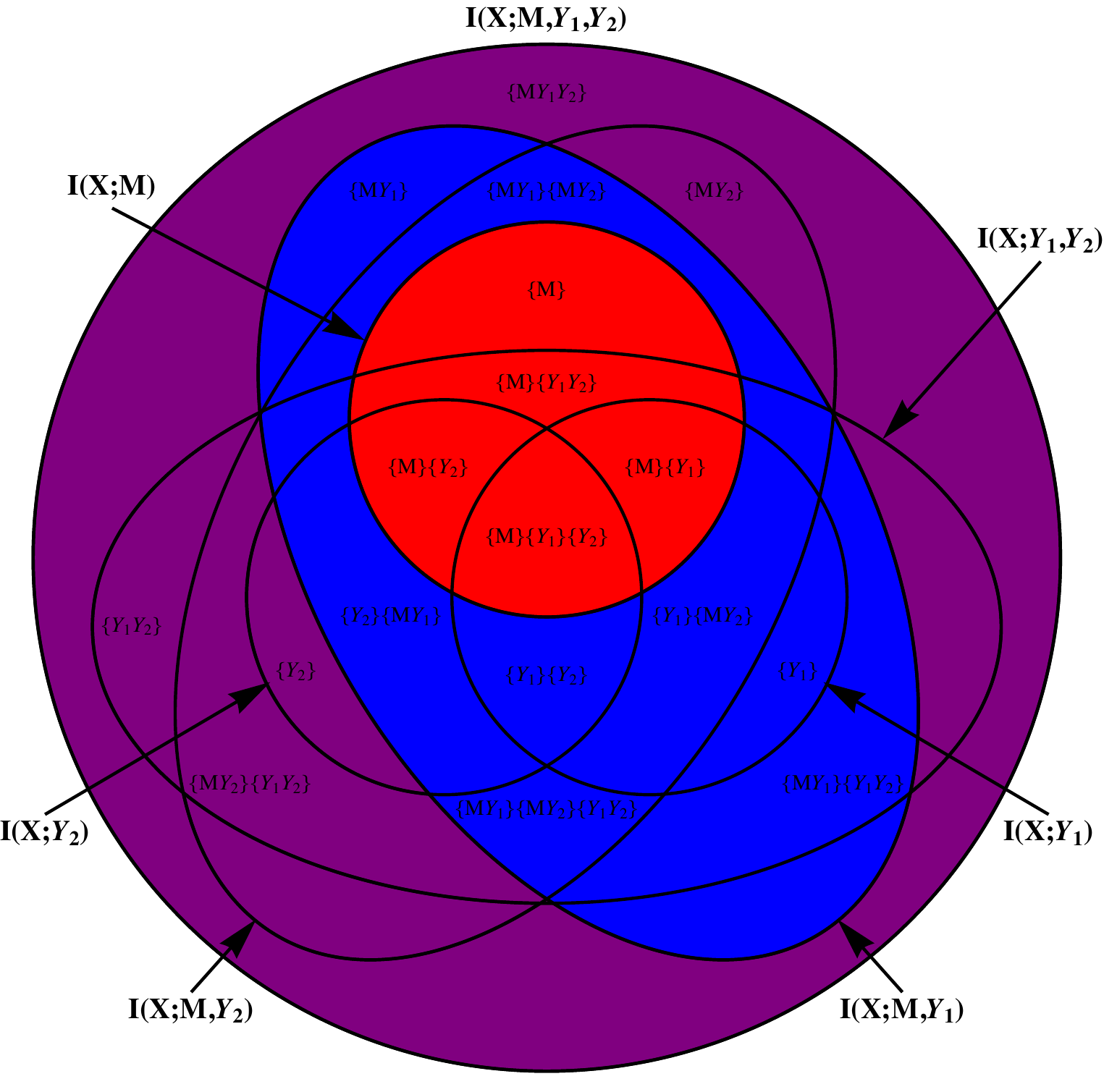}
	\end{center}
	\caption{\label{fig:storageAndTransferPartialDiagram} PI-diagram of information in $X$ decomposed from \emph{three} source variables $M, Y_1, Y_2$ (ignoring the colors). If we consider $M$ to be the past state $\mathbf{X^{(k)}}$ of the destination $X$, and $Y_1$ and $Y_2$ as two other causal sources, then this PI-diagram partitions the AIS (red) and transferred information (all other information; blue and purple here). The transferred information (from two sources) can be further partitioned into apparent TE from $Y_1$ (blue), then complete TE from $Y_2$ (purple). (This is called the PI-diagram for \emph{four} variables in \cite{will10a}, including the destination variable.)}
\end{figure}

Using $I_\cap$ and a form of inclusion-exclusion, the PID framework specifies how the total information $I(X; \mathbf{A})$ decomposes into a sum of PI-terms, given by the function $I_\partial$.
In the simplest case of two source variables, the total information $I(X; A_1, A_2)$ decomposes into: a. the \emph{redundant} information about $X$ which is shared by both $A_1$ and $A_2$: $I_\partial(X; \{A_1\}\{A_2\}) = I_\cap(X; A_1, A_2)$; b. the \emph{unique} information from $A_1$ (resp. $A_2$): $I_\partial(X; \{A_1\}) = I(X; A_1) - I_\cap(X; A_1, A_2)$: ; and c. the \emph{synergistic} information which can only be identified when $A_1$ and $A_2$ are considered jointly as $\{ A_1, A_2 \}$: $I_\partial(X; \{A_1, A_2\}) = I(X; A_1, A_2) - I(X; A_1) - I(X; A_2) + I_\cap(X; A_1, A_2)$.
The relationships between synergy, redundancy, and unique information can be represented using a PI-diagram (see \fig{partial2Source}), which shows the set-theoretic breakdown of $I(X; A_1, A_2)$ into PI-terms.
Without a valid measure for redundancy, it would not be possible to separately measure these four PI-terms using only the three independent standard information-theoretic terms $I(X; A_1, A_2)$, $I(X; A_1)$ and $I(X; A_1)$.
The PI-diagram for three source variables is shown in \fig{storageAndTransferPartialDiagram}, and from this the general structure of PI decomposition can be seen.
In general, the PI-term $I_\partial(X;\alpha)$ for a collection of sources $\alpha$ corresponds to the information provided redundantly by the synergies of all sources in the collection, corresponding to one distinct way for the source variables to contribute information about the destination.  
Put another way, $I_\partial(X;\alpha)$ is ``the information provided redundantly by the sources of $\alpha$ that is not provided by any simpler collection of sources'' \cite{will10a}, where any simpler collection $\beta$ is lower than $\alpha$ on the hierarchy (or \emph{redundancy lattice}) of the set-theoretic breakdown of $I(X;\mathbf{A})$:
\begin{align}
	I_\partial(X; \alpha) = I_\cap(X; \alpha) - \sum_{\beta \prec \alpha}{I_\partial(X; \beta)}
	\label{eq:iPartial}.
\end{align}
The boundary case is for $\alpha$ with no simpler collection of sources, where $I_\partial(X; \alpha)$ is simply the redundancy $I_\cap(X; \alpha)$.

%
%

\subsection{The $I_{\min}$ measure for redundancy}

The abstract formulation of PI decomposition works for any redundancy measure that satisfies the axioms for $I_\cap$.
However, to actually compute PI-terms, a \emph{concrete} redundancy measure satisfying this axiomatic definition is needed.
Williams and Beer proposed the following candidate measure \cite{will10a}:
\begin{eqnarray}
	I_{\min}(X; \mathbf{A}_1, \ldots, \mathbf{A}_r) = \sum_{s}{ p(s) \  \underset{\mathbf{A}_j}{\min} \ I(X = x; \mathbf{A}_j) }
	\label{eq:iMin}, \\
	I(X = x; \mathbf{A}) = \sum_{\mathbf{a}}{p(\mathbf{a}|x) \left[ \log_2{\frac{1}{p(x)}} - \log_2{\frac{1}{p(x|\mathbf{a})}} \right] }
	\nonumber.
\end{eqnarray}
PI-terms $I_\partial(X; \mathbf{A}_1, \ldots, \mathbf{A}_r)$ which are measured using $I_{\min}$ for $I_\cap$ are labeled as $\Pi(X; \mathbf{A}_1, \ldots, \mathbf{A}_r)$.

$I_{\min}$ measures redundancy as the minimum \emph{amount} of information which can be found in any source $\mathbf{A}_j$.
This has been criticized since it does not specifically require each source to hold the \emph{same} information, as demonstrated with the ``two-bit copy problem'' \cite{tim12a,grif12a,hard12a}, which is the observation that $I_{\min}(\{A_1,A_2\};A_1,A_2) = 1$ bit for independent random bits $A_1,A_2$.
This observation prompted the proposal of a new axiom for $I_\cap$ \cite{hard12a}:
\begin{axiom}
\emph{Identity}: $I_\cap(\{A_1,A_2\};A_1,A_2) = I(A_1;A_2)$.
\end{axiom}
Alternatives measures of redundancy which satisfy this additional axiom have been proposed by Harder et al. \cite{hard12a} and Griffith and Koch \cite{grif12a}.
We describe these briefly in \secRef{prospectsWithOthers}, though focus on $I_{\min}$ in our current study as the originally-presented concrete measure.

\subsection{PI-decomposition of information dynamics}
\label{sec:pidOnInfoDynamics}

PID can clearly be applied to the information sources for a destination as defined by information dynamics for distributed computation; i.e. 
the set $\mathbf{S_{DC}}=\{ \mathbf{X^{(k)}}, Y_1, \ldots , Y_g\}$, including 
the previous state of the destination, and the other causal sources.
This is a partitioning of the information in the next state of the destination variable into information storage and complex transfer terms, and their sub-components.
\fig{storageAndTransferPartialDiagram} shows the PI-diagram for these components; the identification of AIS and apparent TE in this PI-diagram was first given in \cite{fleck11a}, and is akin to the decomposition given in \eq{miBreakdown1}.

Considering the apparent TE $T_{Y \rightarrow X}(k)$ as a conditional MI, Williams and Beer \cite{will11a} note that it is composed of a unique component $I_\partial(X';\{Y\})$ from the source $Y$ (\emph{state-independent TE}) plus a synergistic component $I_\partial(X';\{Y,\mathbf{X^{(k)}}\})$ from the source $Y$ interacting with the past state $\mathbf{X^{(k)}}$ (\emph{state-dependent TE}) (see \fig{partial2Source}).
The case for the conditional/complete TE is more complicated again (see \fig{storageAndTransferPartialDiagram}), where there are many more varieties of synergistic components involved.
%
Similarly, Flecker et al. \cite{fleck11a} suggested that breaking down the PI-terms of the storage and transfer measures can reveal further insights into the local dynamics of a system.
(We will revisit the approach to localizing these components in \secRef{initialApproaches}).

Finally, note the \emph{role of the past state} of the destination $\mathbf{X^{(k)}}$ as a joint source here.
Using different values of $k$ changes the values of the PI-terms, redistributing the decomposition of the information amongst them.
(The information attributed to storage in $I(X';\mathbf{X^{(k)}})$ is non-decreasing with larger $k$, which may decrease information in other PI-terms).
For our purposes $k \rightarrow \infty$ should be used, to align with proper measurement of information storage and transfer (as described in \secRef{infoDynamics}).
The use of large $k$ for $\mathbf{X^{(k)}}$ is not about gathering all causal sources in the past of the destination (indeed, it's unlikely that most of these values will be directly causal to $X'$).
It is about providing \emph{context} for our analysis, or providing the \emph{perspective} of distributed computation \cite{liz08a,liz12a} by properly identifying information storage and transfer in the PI-diagram.

\section{Modified and non-modified information}
\label{sec:infoModification}

Given our view in \secRef{intro} of information modification as the \emph{synthesis} of information from more than one information storage or transfer source alone, the PID has an obvious application here.
In this section, we first briefly review recent initial approaches to measuring information modification, before proposing how to properly capture it in the PI-diagram.

\subsection{Initial approaches}
\label{sec:initialApproaches}

The separable information was introduced by Lizier et al. \cite{liz10e} to capture the information gathered by an observer about the next state of $X$ from \emph{separate} inspection of the storage and transfer sources.
Locally, it is defined simply as:
\begin{align}
	s_{X}(n,k) & = a_X(n,k) + \sum_{Y \in \{Y_1,\ldots ,Y_g \} \setminus X }{ t_{Y \rightarrow X}(n, k) }
		\label{eq:localSeparableFiniteK}.
\end{align}
The intuition behind the separable information was that local AIS and TE become negative where unconsidered sources act strongly to create an outcome in the destination.
It was hypothesized that if $s_{X}(n,k) < 0$, then no source provides strong positive information about the outcome when inspected individually and a non-trivial information modification must be taking place. 
Indeed, $s_{X}(n,k)$ was the first method to directly identify particle collisions in CAs as information modification events \cite{liz10e}.
However, it was acknowledged in \cite{liz10e}
that $s_{X}(n,k)$ ignored interaction or redundancies between the sources, and indeed with the mechanics of PID available, Flecker et al. \cite{fleck11a} identified which components in the PI-diagram of \fig{storageAndTransferPartialDiagram} were double-counted and ignored by $s_{X}(n,k)$.
As such, $s_{X}(n,k)$ remains a \emph{heuristic} rather than a measure, though it guides us in the right direction.
It seems that $s_{X}(n,k) < 0$ was a good predictor of modification events because $s_{X}(n,k) < 0$ events are likely to have strong synergistic components in the PI-diagram, and these synergistic components are more likely to measure the information modification.

Building on these insights, Flecker et al. \cite{fleck11a} suggested that a more natural way of ``quantifying the extent to which the whole contributes information beyond the sum of the parts'' for ECAs would be the 3-way synergy $\Pi(X';\{ \mathbf{X^{(k)}},  Y_1^{(k)}, Y_2^{(k)} \})$ between $\mathbf{X^{(k)}}$ and the two neighboring causal sources $Y_1$ and $Y_2$  (akin to the outer-most PI-term in \fig{storageAndTransferPartialDiagram} but with full states of $Y_1$ and $Y_2$ instead of single values).
This generalizes as the highest-order synergy term in the PI-diagram between the storage and transfer sources.
While this is certainly a proper information-theoretic measure,
it did not work as well in identifying particle collisions in complex CA rules \cite{fleck11a}.
A possible factor was the perspective in \cite{fleck11a} that transfer and modification were mutually exclusive concepts.
This would (as discussed later) ignore the state-dependent TE \cite{will11a}, a constituent of information transfer which captures the interaction between the source and the past state of the destination.
This may have led the identified measure to miss some possible contributions to the modification (i.e. lower-order synergy terms).
Furthermore, the localisation of the PI terms in \cite{fleck11a} was a sliding window, which as discussed in \secRef{localising} does not properly attribute a local value to a specific configuration.

\subsection{Requirements for a measure of information modification}
\label{sec:requirementsForInfoMod}

Having evaluated these attempts to measure information modification in distributed computation, we propose the following requirements that a measure of information modification $M_X$ should satisfy. $M_X$ should:
\begin{enumerate}
\item \label{item:properInfoTheQuant} be a proper information-theoretic quantity;
\item \label{item:useStateK} examine the interaction between the information storage $\mathbf{X^{(k)}}$ and causal transfer sources $Y \in \{ Y_1, \ldots , Y_g\}$;
\item \label{item:localisation} allow local measurement $m_X$ at specific observed configurations $\left( x_{n+1}, x_n^{(k)}, y_{1,n}, \ldots , y_{g,n} \right)$ (defined in more detail in \secRef{localising});
\item \label{item:arbitraryNumSources} be extendible to an arbitrary number of sources $g$.
\end{enumerate}

Clearly, the separable information fails to satisfy requirement \ref{item:properInfoTheQuant}, while the 3-way synergy as localized via sliding windows in \cite{fleck11a} does not satisfy requirement \ref{item:localisation}.


Also, we expect that requirement \ref{item:useStateK}, which gives the perspective of distributed computation in using the past state of the destination $\mathbf{X^{(k)}}$, will be important (i.e. using $k=1$ say would not suffice).
This is because we know that measures of information storage and transfer do not properly align with our understanding of these concepts without large $k$ \cite{liz08a,liz12a}, and similarly large $k$ was required for the precursor heuristic separable information to identify collision points in CAs.

\subsection{Partitioning modified and non-modified information}
\label{sec:infoModViaPID}

We return to our accepted definition of information modification as interactions between transmitted and/or stored information which result in a modification of one or the other.
We expect to split the total information $I(X'; \mathbf{S_{DC}})$ about the destination $X'$ from the information sources $\mathbf{S_{DC}} = \{ \mathbf{X^{(k)}}, Y_1, \ldots , Y_g\}$ into modified information $M_X$ and non-modified information $I(X'; \mathbf{S_{DC}}) - M_X$.

As hinted at previously, we identify the \emph{non-modified information} in the destination $X'$ as any information that is identifiable in any one of the information sources in $\mathbf{S_{DC}}$ examined individually.
In terms of PID, this is the sum of all PI-terms which consider collections of joint sources where (at least) one set of joint sources is only a single source:
\begin{align}
	I(X'; \mathbf{S_{DC}}) - M_X = \sum_{\substack{\beta \preceq \mathbf{S_{DC}} \\ \exists \ \gamma \in \beta, \  |\gamma| = 1} }{I_{\partial}(X'; \beta) }
	\label{eq:nonmodified}.
\end{align}

%
\begin{figure}
	\begin{center}
		\includegraphics[width=0.48\textwidth]{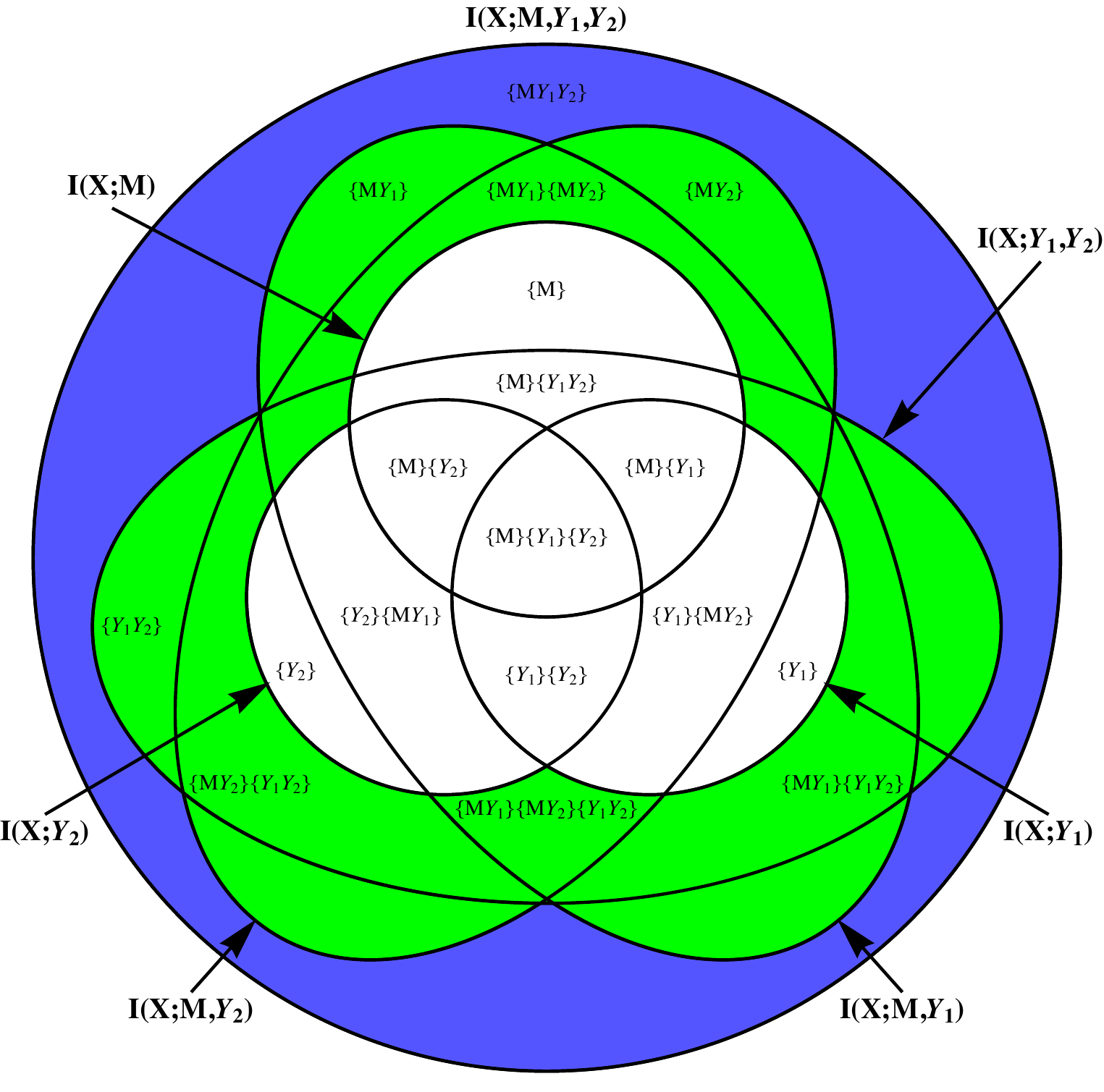}
	\end{center}
	\caption{\label{fig:modifiedKElementsPartialDiagram} PI-diagram of information in a destination $X$ from three source variables $Y_1$, $Y_2$, and $M=\mathbf{X^{(k)}}$, identifying: a. non-modified information (no color), which could be decoded by examining individual sources only, and b. modified information (light-blue and green regions), composed of the information about the destination that could only be decoded by looking at all 3 sources $I_{\partial}^{(o=3)}$ (light-blue), and the information that could be decoded by examining only 2 sources together (but not singles) $I_{\partial}^{(o=2)}$ (green).}
\end{figure}

Conversely then, we can define $M_X$ directly from \eq{nonmodified}.
Equivalently, we can say that the \emph{modified information} $M_X$ is the sum of all synergy terms in the PI-diagram for $I(X'; \mathbf{S_{DC}})$; i.e. all atoms in the PI-diagram which consider collections of joint sources, where no set of joint sources in the collection only considers a single source:
\begin{align}
	M_X = \sum_{\substack{\beta \preceq \mathbf{S_{DC}} \\ \forall \  \gamma \in \beta, \  |\gamma| > 1} }{I_{\partial}(X'; \beta) }
	\label{eq:modified}.
\end{align}
$M_X$ includes any information that cannot be found in one of the sources examined individually, i.e. that which is produced from a non-trivial combination of information from two or more sources in $\mathbf{S_{DC}}$. 
Both modified and non-modified information can be easily identified on the PI-diagram - see \fig{modifiedKElementsPartialDiagram}.

This approach is along the lines suggested in \cite{fleck11a}, but includes more PI-terms.
The key difference is that we include any PI-terms whose collections of joint variables contain at minimum two variables; as such, this measure includes \emph{all} synergistic information terms.\footnote{Also, including spurious uncorrelated sources in addition to $\{ Y_1, \ldots , Y_g\}$ will remove all information in the highest-order synergy term used in \cite{fleck11a}, yet $M_X$ remains the same since it still counts all synergistic PI-terms.
}
In comparison to \cite{fleck11a}, we do not consider the concepts of information transfer and modification to be mutually exclusive.
As shown by the decomposition in \fig{storageAndTransferPartialDiagram}, all of the information in the destination $X$ is either stored information from its past, or (some type of) transferred information from the other sources.
Our view is that modified information is simply the synergistic parts of such information transfer.
To clarify this point with a more simple example, consider the two ``source'' PI-diagram in \fig{partial2Source}.
Here, our approach would label the synergy term $I_{\partial}(X'; \{ \mathbf{X^{(k)}}, Y \})$ as the information modification $M_X$, and note that in this case the quantity is precisely equal to the state-dependent TE, which is a constituent of information transfer \cite{will11a}.

\subsection{Hierarchy of orders of interaction}
\label{sec:hierarchies}

We can also define a \emph{hierarchy} of the decomposition, in terms of the \emph{minimum number} of interacting joint sources that information about the destination could be found in.
For a generic PI-diagram with sources $\{ \mathbf{A}_1, \ldots , \mathbf{A}_r \}$, the information which could be decoded from only $o$ sources but not $o-1$ sources is:
\begin{align}
	I_{\partial}^{(o)}(X'; \{ \mathbf{A}_1, \ldots , \mathbf{A}_r \}) & = \sum_{\substack{\beta \preceq \{ \mathbf{A}_1, \ldots , \mathbf{A}_r \} \\ \min(| \gamma |) = o, \  \gamma \in \beta} }{I_{\partial}(X'; \beta) }
	\label{eq:hierarchyOfInteraction1}, \\
	I(X'; \{ \mathbf{A}_1, \ldots , \mathbf{A}_r \}) & = \sum_{o=1}^{r}{I_{\partial}^{(o)}(X'; \{ \mathbf{A}_1, \ldots , \mathbf{A}_r \}) }
	\nonumber 
\end{align}
We note that this addresses the goal of \cite{kah09,lud08}, to achieve a partitioning of information in a given variable or collective into a hierarchy of contributions from individual sources, from pairs of sources that was not contained in individuals, etc. In comparison to these approaches however, $I_{\partial}^{(o)}$ avoids problematic double-counting and the use of the negative ``interaction information'' \cite{will10a} (unlike \cite{lud08}), and (depending on the concrete implementation of $I_{\partial}$) is model-free (unlike \cite{kah09}).

Using the distributed computation sources $\mathbf{S_{DC}}$, we have:
\begin{align}
	M_X & = I(X'; \mathbf{S_{DC}}) - I_{\partial}^{(o=1)}(X'; \mathbf{S_{DC}})
	\label{eq:modifiedInTermsOfHierarchy1}, \\
	M_X & = \sum_{o=2}^{g+1}{ I_{\partial}^{(o)}(X'; \mathbf{S_{DC}}) }
	\label{eq:modifiedInTermsOfHierarchy2},
\end{align}
and clearly for the three source case $\{ \mathbf{X^{(k)}}, Y_1, Y_2 \}$ in \fig{modifiedKElementsPartialDiagram} we have $M_X = I_{\partial}^{(o=2)} + I_{\partial}^{(o=3)}$.

\subsection{Modified information in ECAs}
\label{sec:averageModifiedInfoInECAs}

We apply our definition of modified information to several important ECA rules, using the $I_{\min}$ candidate redundancy measure and $\Pi$ to compute $M_X^\Pi$ (as implemented in the publicly available software \cite{liz12e}).
Our results in \tableRef{caResults} show that for simple, ordered CA rules, non-modified information dominates the decomposition of the next state of a cell.
Conversely, for chaotic CAs (rules 18, 22 and 30), modified information dominates, resulting from synergistic interactions between sources.
The complex CAs (rules 54 and 110) however seem to have a mix of modified and non-modified information.
These results make intuitive sense, and align with previous observations in both CAs and random Boolean networks that chaotic dynamics tend to be dominated by higher-order information transfer terms \cite{liz08a,liz10e,liz11b}.

%
%
\begin{table}[!t]
\caption{Measurements (in bits) to 3 d.p. of the hierarchies of modified and non-modified information in ECAs, using the $I_{\min}$ redundancy measure.
We use observations of 100 repeat runs of length 200 CAs run for 200 time steps,
with history length $k=16$ here except for $k=1$ in the last column.}
\label{table:caResults}
\centering
\begin{tabular}{|c|c|c|c|c|c|}
 \hline
 Rule & $\Pi^{(o=1)}$ & $\Pi^{(o=2)}$ & $\Pi^{(o=3)}$ & $M_X^\Pi(k=16)$ & $M_X^\Pi(k=1)$\\
 \hline
 18  & 0.273 & 0.464 & 0.087 & \textbf{0.551} & 0.691 \\
 22  & 0.188 & 0.188 & 0.559 & \textbf{0.747} & 0.916 \\
 30  & 0.189 & 0.558 & 0.253 & \textbf{0.811} & 0.812 \\
 54  & 0.705 & 0.087 & 0.205 & \textbf{0.292} & 0.860 \\
 110 & 0.689 & 0.177 & 0.121 & \textbf{0.298} & 0.899 \\
 \hline
\end{tabular}
\end{table}

The same analysis run with only $k=1$ past value for $\mathbf{X^{(k)}}$ does not provide the same insight, in fact identifying large amounts of information from triplet interactions for all the rules.
This is because using $k=1$ does not adequately partition information storage and transfer, and so does not achieve a proper perspective of distributed computation (as expected from \secRef{requirementsForInfoMod}).

We would like to evaluate the dynamics of information modification in space and time -- in the same manner as shown for AIS in \fig{caLocalResults} -- since this will reveal whether they relate to particle collisions in CAs.
To do so, we require the ability to compute the value of PI-terms on a local rather than average scale, and we consider this in the next section.

\section{Localising PI-terms}
\label{sec:localising}

The ability to localize PI-terms depends on the ability to localize the measure of redundancy $I_\cap$ to obtain relevant local values $i_\cap$.
Local PI-terms $i_\partial$ would be the sums of the relevant $i_\cap$, as per the standard values.
However, a property of localizability of the abstract measure $I_\cap$ does not follow from its definition by the original minimal set of axioms in \cite{will10a}, and so at this stage the localizability will be a property of the concrete measure (e.g. $I_{\min}$) one selects to implement $I_\cap$.
Here we consider how one may define localizability of $I_\cap$ in terms of a further axiom, and subsequently consider whether the candidate concrete measures satisfy these axioms.

\subsection{Localizing redundancy $I_\cap$}

For a candidate redundancy measure to be localizable (as defined for traditional measures in \secRef{infotheory}), it must satisfy the following additional axiom for $I_\cap(X; \mathbf{A}_1, \ldots, \mathbf{A}_r)$ :
\begin{axiom}
\emph{(localizability)} There exists a local measure $i_\cap(x; \mathbf{a}_1, \ldots, \mathbf{a}_r)$ for the redundancy of a specific observation $\{ x, \mathbf{a}_1, \ldots, \mathbf{a}_r \}$ of $\{ X, \mathbf{A}_1, \ldots, \mathbf{A}_r \}$, such that:
\begin{enumerate}
\item $i_\cap(x; \mathbf{a}_1, \ldots, \mathbf{a}_r)$ satisfies the corresponding symmetry and self-redundancy axioms as per $I_\cap(X; \mathbf{A}_1, \ldots, \mathbf{A}_r)$;
\item $I_\cap(X; \mathbf{A}_1, \ldots, \mathbf{A}_r) = \left\langle i_\cap(x; \mathbf{a}_1, \ldots, \mathbf{a}_r) \right\rangle$;
\item $i_\cap(x; \mathbf{a}_1, \ldots, \mathbf{a}_r)$ is once-differentiable with respect to changes in $p(x, \mathbf{a}_1, \ldots, \mathbf{a}_r)$; and
\item $i_\cap(x; \mathbf{a}_1, \ldots, \mathbf{a}_r)$ is uniquely defined for the given candidate redundancy measure.
\end{enumerate}
\end{axiom}

Note that the self-redundancy axiom here means that $i_\cap(x; \mathbf{a}) = i(x; \mathbf{a})$; i.e. local self-redundancy is simply a local MI.
As such, the relevant local MI terms should be sums of the relevant local PI-terms $i_\partial$.
We recall that local MI terms are unique, symmetric, and additive, whilst averaging to give the relevant MI, and are once-differentiable with respect to small changes in the PDFs \cite{fano61}, and the above axiom requires several similar features.
Now, there is no requirement for the \emph{local} values $i_\cap$ to satisfy monotonicity (unlike the average), in a similar way to local MI values being able to increase or decrease with the number of variables so long as the average MI increases.
Similarly, since local MI values can be negative, then local redundancy and PI-terms may also be negative.

Sliding window methods are not local values, since they do not provide a value for a specific configuration (but are a function of the window as a whole). As such, the approach used in \cite{fleck11a} is not an appropriate localization.

With regard to continuity of $i_\cap(x; \mathbf{a}_1, \ldots, \mathbf{a}_r)$, we note from an information geometry perspective, the local value is effectively a function of $d$ variables, where $d$ is the number of degrees of freedom in defining $p(x, \mathbf{a}_1, \ldots, \mathbf{a}_r)$ in the space of such probability distributions.
The continuity of $i_\cap(x; \mathbf{a}_1, \ldots, \mathbf{a}_r)$ can be thought of as being with respect to these variables defining $p(x, \mathbf{a}_1, \ldots, \mathbf{a}_r)$.
Notably, Shannon required such continuity in defining the entropy \cite{shan48}.

Uniqueness of $i_\cap(x; \mathbf{a}_1, \ldots, \mathbf{a}_r)$ will depend on the specific definition of the concrete redundancy measure.

Finally, we argue that the motivation for a redundancy measure to satisfy localizability goes well beyond our desire to measure information modification on a local scale.
This property would make the dynamics of \emph{any} PI-term measurable on a local scale in space and time, as for other measures.


\subsection{Localising $I_{\min}$}
\label{sec:cantLocaliseImin}

The straightforward way to localize $I_{\min}$ for a specific observation $\{ x, \mathbf{a}_1, \ldots, \mathbf{a}_r \}$ of $\{ X, \mathbf{A}_1, \ldots, \mathbf{A}_r \}$ is to take:
\begin{align}
	i_{\min}(x; \mathbf{a}_1, \ldots , \mathbf{a}_r) & = i(x;\mathbf{a}_j)
    = \log_2{\frac{p(x \mid \mathbf{a}_j)}{p(x)}}
	\label{eq:iminlocal2},
\end{align}
where $\mathbf{a}_j$ is the specific value of $\mathbf{A}_j$ in this observation where:
\begin{align}
	\mathbf{A}_j = \underset{\mathbf{A}_j}{\argmin} \  I(X = x; \mathbf{A}_j)
	\label{eq:argminOfPartialMi}.
\end{align}
Recalling that $I_{\min}$ is the ``minimum information that any source provides about each outcome'' of the destination variable ``averaged over all possible outcomes'' \cite{will10a}, here $i_{\min}$ is the information provided about the destination observation by the specific observation of the source $\mathbf{A}_j$ which provides the minimum information on average.
This localization averages directly over $p(x)p(\mathbf{a}_j|x)$ (as per \eq{iMin}) to give $I_{\min}(X; \mathbf{A}_1, \ldots, \mathbf{A}_r)$, and at first seems to satisfy our axiom.

However, it is simple to demonstrate that $i_{\min}(x; \mathbf{a}_1, \ldots , \mathbf{a}_r)$ is not once-differentiable with respect to changes in the PDF $p(x, \mathbf{a}_1, \ldots , \mathbf{a}_r)$.
Let us take the Boolean OR function for binary variables, $X = A_1 + A_2$, and assume that we have an almost equiprobable distribution of the inputs $(A_1,A_2)$ as shown in \tableRef{orFunction}.
A small disturbance $\delta \rightarrow 0^+$ to the equiprobable distribution is enough to ensure that $A_j=A_1$ is always selected by the $\min$ function here\footnote{For $X=0$, since $A_1 = 0$ (slightly) more often when $X \neq 0$, then $A_1$ tells us less specific information about $X$. Similarly, $A_1 = 1$ (slightly) less often when $X=1$, so again tells us less specific information about $X$.}, giving the local values for redundancy $i_{\min}(x; \{a_1\},\{a_2\})$ displayed in \tableRef{orFunction}.
If the infinitesimal disturbance $\delta$ changes sign however (causing a continuous change in the underlying PDF $p(x,a_1,a_2)$), this flips the selection of $A_j$ to $A_2$, and \emph{discontinuously} swaps the local values of $i_{\min}(x; \{a_1=0\},\{a_2=1\})$ and $i_{\min}(x; \{a_1=1\},\{a_2=0\})$.
Also, with $\delta=0$ there are two possible solutions for the local values, meaning the uniqueness requirement is not satisfied either.
As such, this localization for $I_{\min}$ does not satisfy the localizability axiom.

\begin{table}[!t]
\caption{Redundancy $\pi(x; \{a_1\},\{a_2\}) = i_{\min}(x; \{a_1\},\{a_2\}) = i(x;a_j)$ for the OR function $X = A_1 + A_2$, with an equiprobable input distribution slightly disturbed by an infinitesimal $\delta \rightarrow 0^+$.}
\label{table:orFunction}
\centering
\begin{tabular}{|c|c|c|c|c|}
 \hline
 $a_1,a_2$ & $x$ & $p(a_1,a_2)$ & $\underset{A_j}{\argmin} \  I(X = x; A_j)$ & $i(x;a_j)$ \\
 \hline
 0,0 & 0 & 0.25 & $A_1$ & 1\\
 0,1 & 1 & $0.25 + \delta$ & $A_1$ & -0.585\\
 1,0 & 1 & $0.25 - \delta$ & $A_1$ & 0.415\\
 1,1 & 1 & 0.25 & $A_1$ & 0.415 \\
 \hline
\end{tabular}
\end{table}

It is tempting to define $i_{\min}$ as the minimum information that any specific source observation provides about the destination observation (i.e. taking the $\min$ of local values $i(x;\mathbf{a}_j)$), however this would not average over all observations to give $I_{\min}$.
Aside from this, at this stage there are no other clear meaningful candidates for localization of $I_{\min}$.


\subsection{Prospects with other candidate redundancy measures}
\label{sec:prospectsWithOthers}

There is the prospect that alternate measures satisfying the existing axioms for $I_\cap$ may satisfy the axioms we have laid out above for localizing redundancy and information modification.
Two candidates here \cite{grif12a,hard12a} were proposed to address the two-bit copy problem raised with $I_{\min}$.

Griffith and Koch propose to measure the redundancy by mapping the destination $X$ to a surrogate $X'$ which preserves the information from each source $\mathbf{A}_j$ to the surrogate, but minimizes the overall mutual information from the sources to the surrogate \cite{grif12a}.
This method at first seems localizable (by simply localizing the MI between the sources and the surrogate), however as pointed out in \cite{grif12a} the mapping (i.e. PDFs) to produce the minimal MI is not unique.
As such, the method does not immediately satisfy the uniqueness requirement for our localizability axiom, though potentially extra conditions could be added to the definition in future to meaningfully uniquely identify the minimizing mapping.

Harder et al. \cite{hard12a} propose an information geometry based approach.
This involves projecting the conditional distributions of the destination $X$ given each source $\mathbf{A}_j$ onto eachother in the relevant information-geometric space.
At first glance this method seems localizable.
However, it is not currently suitable for our purposes in investigating information modification, since it is currently only defined for a pair of sources.
If it can be extended to an arbitrary number of sources, it should satisfy our requirement \ref{item:arbitraryNumSources} in \secRef{requirementsForInfoMod} for applicability to capture information modification via a PI-diagram.

\section{Conclusion}
\label{sec:conclusion}

We have described how frameworks for information dynamics and partial information decomposition could be used together to describe the modification of information in distributed computation.
This involves examining the partial information diagram for the information storage and transfer sources to a destination, and then identifying synergies for pair interactions and above as information modification.

We applied the $I_{\min}$ measure of redundancy to cellular automata in this fashion, and demonstrated that ordered CAs have little modified information, the dynamics of chaotic CAs are dominated by information modification, while complex CAs have an intermediate level.
It remains to be seen whether the overall nature of these results would change if using an alternative redundancy measure to $I_{\min}$ (e.g. \cite{grif12a,hard12a}).

Examining the dynamics of such information modification on a local scale in space and time requires localizability of the given redundancy measure that one uses to compute the PI-terms.
We have suggested an axiom that such a measure should satisfy for it to be localizable, and demonstrated that the $I_{\min}$ measure does not satisfy this axiom.
%
Finally, we assessed the potential for other candidate redundancy measures to be applied to local information modification. 
We found that none are suitable in their current form, but there is potential for them to be extended to meet our requirements.

\section*{Acknowledgment}

PLW was supported by NSF grants IIS-0916409 and  IIC-1216739.

\bibliographystyle{IEEEtran}
\bibliography{refs}

\end{document}